\documentclass[useAMS, usegraphicx, usenatbib]{mn2e}
 \usepackage{amsmath, amssymb, mathrsfs, dcolumn,mathptmx,aas_macros}
\newcommand{\D}{\tilde{\nabla}}
\newcommand{\curl}{\mathrm{curl}}
\newcommand{\bq}{\begin{equation}}
\newcommand{\eq}{\end{equation}\noindent}
\newcommand{\bqa}{\begin{eqnarray}}
\newcommand{\eqa}{\end{eqnarray}\noindent}
\newcommand{\bsub}{\begin{subequations}}
\newcommand{\esub}{\end{subequations}}

\begin{document}

\title[Scalar Perturbations in $2$-Temperature Plasmas]{Scalar Perturbations in Two-Temperature Cosmological Plasmas}

\author[J.~Moortgat and M.~Marklund]{J.~Moortgat$^{1}$\thanks{Email: moortgat@pas.rochester.edu. Current address: Department of Physics \& Astronomy, University of Rochester, Bausch \& Lomb Hall, P.O. Box 270171, 600 Wilson Boulevard, Rochester, NY 14627-0171.} and M.~Marklund$^{2}$\thanks{Email: mattias.marklund@physics.umu.se.}\\
$^{1}$ Department of Astrophysics, {\sc imapp}, Radboud University, PO Box 9010, 6500 GL Nijmegen, The Netherlands.\\
$^{2}$ Centre for Nonlinear Physics, Department of Physics, Ume{\aa} University, SE--901 87 Ume{\aa}, Sweden.}

\maketitle

\begin{abstract}
We study the properties of density perturbations of a two-component plasma with a temperature difference on a homogeneous and isotropic background. For this purpose we extend the general relativistic gauge invariant covariant ({\sc gic}) perturbation theory to include a multi-fluid with a particular equations of state (ideal gas) and imperfect fluid terms due to the relative energy flux between the two species. We derive closed sets of {\sc gic} {\em vector} and subsequently {\em scalar} evolution equations. 
We then investigate solutions in different regimes of interest. In particular, we study long wavelength and arbitrary wavelength Langmuir and ion-acoustic perturbations. The harmonic oscillations are superposed on a Jeans type instability. We find a generalised Jeans criterion for collapse in a two-temperature plasma, which states that the species with the largest sound velocity determines the Jeans wavelength. Furthermore, we find that within the limit for gravitational collapse, initial perturbations in either the total density or charge density lead to a growth in the initial temperature difference.
These results are relevant for the basic understanding of 
the evolution of inhomogeneities in cosmological models. 
\end{abstract}
%\pacs{52.27.Ny, 04.40.-b, 98.80.-k}

%-----------------------------------------
\section{Introduction}
%-----------------------------------------
Plasmas and electromagnetic fields are common in our Universe.
They play an important role in a diverse setting of astrophysical and cosmological
processes. Plasmas may be found, e.g.~in
stars, accretion disks of rotating black holes, the Earth's
ionosphere, and also constitute the intergalactic, interstellar as well as the
intrasolar medium. There also are occasions when general relativistic gravity has to be taken into account in conjunction with plasma physics, such as in the close vicinity of the aforementioned rotating black holes. Another prominent example is
our Universe, in which the plasma state has over time been more or less prominent. 
Obviously, in such situations gravitational effects due to general relativity, such as gravitational waves, can not be
neglected, and may lead to interesting results.

Moreover, perturbation theory within gravitational physics has always held a 
special place within physics, due to, e.g.\ the pioneering work of Jeans,
and its relation to the basic question of structure formation. The issue 
of structure formation is now a mature science and cosmological observations have been taken into the high precision regime by observational tools such 
as {\sc cobe}, {\sc wmap}, Chandra, and the planned {\sc lisa} mission
The development of new observational instruments also calls for new
theoretical models and concepts to be tested, in order to refine the current 
standard model of cosmology. Thus, there is ample interest in extending the 
existing models by including other physical effects.

In perturbation theory there are two main distinct schools of
thought. The first could
be described as {\em metric based} and follows from a seminal paper by \citet{bardeen} which was extended to multi-fluids by \citet{kodama} and developer further in a more recent paper by \citet{malik}. The second paradigm is a {\em covariant and gauge-invariant perturbation theory} due to \cite{hawking66} and \cite{ellis-bruni} (for reviews see \cite{ellis95} or \cite{Ellis-vanElst}), which was extended to multifluids in \cite{dunsby-bruni-ellis, marklund00, marklund03, betschart04} and applied to the cosmic microwave background ({\sc cmb}) by \citet{challinor}.  The equivalence of the two theories has been shown in \cite{dunsby-bruni-ellis}. 

We will adopt the second approach because the physical quantities defined as gauge invariant and covariant ({\sc gic}) variables are not only convenient mathematically, because they avoid spurious non-physical gauge modes, but also clearly reflect the physical quantities that an observer would measure and these variables allow a clear geometrical interpretation.
In this paper we will apply the theory to a two-component, two-temperature plasma in a cosmological setting. In particular, we assume that the two plasmas each satisfy the perfect gas law, but with different temperatures. We formulate the set of governing equations for the plasma dynamics on a general relativistic background. These equations are analysed by perturbing the two-fluid model around an isotropic and homogeneous background using the covariant gauge invariant approach.

A large-scale temperature difference between the electron and proton fluids in a cosmological setting is not very likely since the Coulomb equilibration time is usually much shorter than the growth rate of perturbations. However, local fluctuations might occur and more importantly we hope to apply the generalized two-fluid equations derived in this paper to other two-temperature fluids in future papers, such as dark matter -- ordinary matter, decoupled free-streaming neutrinos in the early universe and possibly the advection dominated accretion flows onto black holes, where the electrons and protons are thought to be out of thermal equilibrium.

%-----------------------------------------
\section{Gauge invariant covariant theory}
%-----------------------------------------
To make this paper more self-consistent we will recapitulate some of the results in the literature that we use here and make the appropriate approximations to linear perturbations.

\subsection{$3+1$ split}\label{eq::driepluseen}
To start with, we will consider two perfect fluid species that flow in a curved space-time. Both in an astrophysical and a cosmological context it is often possible to identify a preferred family (congruence) of {\em fundamental world lines} associated with the motion of typical observers. In the former case, one could for example choose the world lines of observers that move with the bulk velocity of the matter in a (relativistic) plasma wind or jet associated with a pulsar or active galactic nucleus. In cosmology it is customary to treat the Universe as consisting of a perfect fluid whose world lines are determined by the motion of distance clusters of galaxies with respect to an observer. 

When such fundamental word lines can be identified, it is possible and convenient to split $4$-dimensional space-time into `time' and `space' again with respect to the $4$-velocity vector tangent to the world lines
\bq
u^{a} = \frac{d x^{a}}{d\tau}\quad\Rightarrow\quad u^{a}u_{a}=-1\ ,
\eq
where $\tau$ is the proper time along the world lines. The tensors that project onto $u^{a}$ and into the tangent $3$-spaces orthogonal to $u^{a}$ are given, respectively, by
\bq\label{eq::projections}
U^a_{\ b} \equiv - u^a u_{b} \quad \mbox{and} \quad h_{ab} \equiv g_{ab} + u_{a} u_{b}\ .
\eq
Associated with the $3$+$1$ split are two derivatives. The `time derivative' is the covariant derivative along the fundamental world-lines and the `spatial derivative' is the covariant derivative of any tensor, where all the indices are projected on the hypersurface orthogonal to $u^{a}$. For an arbitrary tensor $T^{ab}_{\ \ cd}$ these are defined by
\bsub
\bqa
\dot{T}^{ab}_{\quad cd} &=& u^{e}\nabla_{e}T^{ab}_{\quad cd}\ ,\\
\D_{e} T^{ab}_{\quad cd} &=& h^{a}_{\ f} h^{b}_{\ g} h^{p}_{\ c}h^{q}_{ \ d}h^{r}_{\ e}\nabla_{r}T^{fg}_{\quad pq}\ .
\eqa
\esub
The derivative $\D$ is a proper $3$-dimensional derivative if and only if the vorticity of $u^{a}$ is zero, for instance in a Friedman-Lema{\^i}tre-Robertson-Walker ({\sc flrw}) Universe (or a {\sc frw} Universe when we neglect the cosmological constant).

\subsection{Multi-fluid stress-energy tensor}
We assume a congruence $u^{a}$ corresponding to fundamental observers as defined in Section~\ref{eq::driepluseen} and a Universe filled with a plasma made up of two species that are allowed to move in arbitrary directions with respect to the observer with individual (non-relativistic) $4$-velocities satisfying
\bq\label{eq::velocity}
u^{a}_{(i)} = u^{a} + v^{a}_{(i)}\ .
\eq 
The index $i=1, 2$ or $i=+, -$ labels the two fluids.
Each species is assumed to be a {\em perfect fluid} in its own rest frame. Then $u^{a}_{(i)}$ is the {\em unique} hydrodynamical $4$-velocity that is time-like and allows a split of the energy-momentum tensor in the perfect fluid form\footnote{This is the $u_{E}^{a}$ $4$-velocity in \cite{dunsby-bruni-ellis}.}
\begin{equation}\label{eq::energymomentum1}
T_{(i)}^{ab} = \mu_{(i)} u_{(i)}^au_{(i)}^b +p_{(i)} h^{ab}\ .
\end{equation}
Here $p_{(i)}$ is the pressure and $\mu_{(i)}$ the energy density and for a perfect fluid $u_{(i)}^{a}$ 
is also parallel to the particle and entropy flux, such that 
\bq
N^{a}_{(i)} = n_{(i)} u_{(i)}^{a}\ , \quad S_{(i)}^{a}=s_{(i)} u_{(i)}^{a}\ ,
\eq
and there is neither a particle drift nor an energy flux ($J_{(i)}^{a} = q_{(i)}^{a} = 0$, $n_{(i)}$ the number density and $s_{(i)}$ the entropy).

In the frame of the fundamental observers $u^{a}$, which in general will move relative to the rest frame of the individual fluids, the energy-momentum tensor is clearly different from \eqref{eq::energymomentum1}. In particular it can be written in the general form of an imperfect fluid. 
Eq.~\eqref{eq::velocity} defines the {\em velocity} $v^{a}_{(i)}$ of the fluid $i$ with respect to the observer $u^{a}$. If we assume that in the background all species share the same hydrodynamical $4$-velocity $u^{a}$ then $v^{a}_{(i)} = 0$ in the background and is therefore gauge invariant by the lemma of \citet{stewartlemma}.

We linearize the stress-energy tensor for non-relativistic deviations from the average flow such that $v^{a}_{(i)}\ll 1$ and the an-isotropic pressure is negligible, but to linear order there is a contribution to the energy flux, or heat, given by $q^{a}_{(i)}\equiv (\mu_{(i)} + p_{(i)})v^{a}_{(i)}$ such that
\bq
T^{ab}_{(i)} = \mu_{(i)} u^{a}u^{b} + p_{(i)} h^{ab} + 2 u^{(a}q^{b)}_{(i)}\ ,
\eq
which are the linearized equivalents of Eqs.~(3-4) in \cite{marklund03}.

\subsection{Equation of state}
In this paper we will use a particular equation of state to study temperature effects in a two-component plasma by a first order perturbation analysis on a {\sc frw} background. We assume a ideal gas equation of state for each species and corresponding energy density and enthalpy $h_{(i)}$:
\begin{subequations}
\bqa\label{eq::eos}
  p_{(i)} &=& k_B n_{(i)} T_{(i)}\ ,\\\label{eq:eqstate}
  \mu_{(i)} &=& \left(m_{(i)} + \tfrac{3}{2}k_B T_{(i)}\right) n_{(i)}\ ,\\
  h_{(i)} &=&  \mu_{(i)} +  p_{(i)} = \left(m_{(i)} + \tfrac{5}{2}k_B T_{(i)}\right) n_{(i)}\ ,
\eqa
\end{subequations}
where $T_{(i)}$ is the temperature,
$m_{(i)}$ the rest mass, and $k_B$ Boltzmann's constant. We will allow for
different temperatures between the species, and note that $T_{(i)}$ is time-dependent
in the background {\sc frw} space-time. Moreover, we will re-scale $k_BT_{(i)} \rightarrow T_{(i)}$ for the sake of brevity. 

\subsection{Conservation of particles, energy and momentum}

Conservation of energy and momentum is guaranteed automatically by the twice-contracted Bianchi identities and Einstein's field equations and is given for each fluid by 
\bq\label{eq::tenscons}
\nabla_{b} T^{ab}_{(i)} = \mu_{0}F^{a}_{\ b}j^{b}_{(i)}\ ,
\eq
where $ T^{ab}_{(i)}$ is the stress-energy tensor for the matter and the right-hand-side follows from the electromagnetic part by $\nabla_{b} T^{ab}_{\mathrm{em},(i)} = - \mu_{0}F^{a}_{\ b}j^{b}_{(i)}$ with $j^{b}_{(i)}$ the $4$-current density and $\mu_{0}$ the vacuum magnetic permeability. Similarly, conservation of number density of particles is given by $\nabla_{a} N_{(i)}^{a}=0$.

The time-like and space-like projections \eqref{eq::projections} of \eqref{eq::tenscons} provide separate linearized equations for energy and momentum conservation, respectively (see for instance \cite{marklund00})
\bsub\label{eq::dtmunu}
\begin{align}\label{eq::muj}
\dot{\mu}_{(i)} &=- h_{(i)}(\Theta + \D_{a} v_{(i)}^{a})\ ,\\\label{eq::vj}
h_{(i)} (\dot{u}^a + \dot{v}^{\langle a\rangle}_{(i)}) &=
  -\D^ap_{(i)} - (\tfrac{\Theta }{3}h_{(i)}  +
   \dot{p}_{(i)}) v^a_{(i)} + \rho_{(i)}E^a\ ,
\end{align}
\esub
where $\Theta$ is the expansion, $\rho_{(i)} = q_{(i)}n_{(i)}$ is the charge density and $E^a$ the electric field strength. 

A separate evolution equation for the temperature can be found from combining \eqref{eq:eqstate} and \eqref{eq::muj} with the linearized equation of particle conservation:
\bsub\label{eq::njTj}
\begin{eqnarray}
  && \dot{n}_{(i)} = -(\Theta + \D_av^a_{(i)})n_{(i)}\ , \\
  && \dot{T}_{(i)} = -\tfrac{2}{3}(\Theta + \D_av^a_{(i)})T_{(i)}\ . 
\end{eqnarray}
\esub

\subsection{Expansion}
The basic equation for gravitational attraction is the {\em Raychaudhuri} equation that in a {\sc frw} Universe with vanishing acceleration, shear, vorticity and cosmological constant\footnote{All equations can easily be generalised to include a cosmological constant, but for simplicity we will specify to a {\sc frw} Universe.} reduces to
\bq\label{eq::thetadot}
\dot{\Theta} = -\frac{\Theta^{2}}{3} - \frac{\kappa}{2} (\mu + 3 p)\ ,
\eq 
with $\kappa = 8\pi G/c^{4}$. \citet{Ellis-vanElst} call the term $\mu + 3 p$ in \eqref{eq::thetadot}  the {\em active gravitational mass density} of the plasma to emphasise that this term and not just $\mu$ drives gravitational contraction. 
The volume rate of expansion $\Theta = \D_{a} u^{a}$ is the trace of the covariant derivative of $u^{a}$ and is related to the Hubble parameter $H$ and the scale-factor $S$ by 
\bq\label{eq::scale}
\frac{\dot{S}(t)}{S(t)} = \frac{\Theta}{3} = H\ .
\eq

\subsection{Electromagnetic field equations}
The electromagnetic field is determined by the linearized Maxwell evolution and constraint equations:
\bsub
\begin{eqnarray}
  && \dot{E}^{\langle a\rangle} = -\tfrac{2}{3}\Theta E^a + \curl\,B^a - \mu_0j^a\ , \\\label{eq::dotB}
  && \dot{B}^{\langle a\rangle} = -\tfrac{2}{3}\Theta B^a - \curl\,E^a\ , \\\label{eq:divE}
  && \D_aE^a = \frac{\rho}{\epsilon_0}\ , \\
  && \D_aB^a = 0\ ,
\end{eqnarray}
\esub
where $\rho = q_{(1)}n_{(1)} + q_{(2)}n_{(2)}$ and $j^a = q_{(1)}n_{(1)}v^a_{(1)} + q_{(2)}n_{(2)}v^a_{(2)}$ are the total 
charge density and total $3$-current density, respectively, $q_{(i)}$ is the particle charge and $\epsilon_{0}$ the vacuum electric permittivity. 

To be more specific, we will choose species $1$ to correspond to protons or positrons with $q_{(1)}=e$ and species $2$ to electrons $q_{(2)}=-e$.

\subsection{Definition of gauge invariant quantities}
Since the background space-time is
{\em homogeneous and isotropic}, we require that the electromagnetic fields vanish to 
zeroth order. This implies that the charge density and current density should vanish and thus $\rho = 0 \Rightarrow n_{(1)} = n_{(2)}$, and $j^{a} = e n (v_{(1)}^{a} - v_{(2)}^{a}) = 0 \Rightarrow v_{(1)}^{a}
= v_{(2)}^{a}$. Because the energy flux should also vanish in the background $q^a = \sum_{(i)}(\mu_{(i)} + p_{(i)})v^a_{(i)} = 0 \Rightarrow v_{(1)}^{a}
= v_{(2)}^{a} = 0$.

Thus, the only nonzero variables in the background are the expansion $\Theta$,
the energy density $\mu_{(i)}$, the number density $N_{(i)}$, and the pressure $p_{(i)}$, or equivalently the temperature $T_{(i)}$. 
The background evolution of these quantities follows from the zeroth order terms in \eqref{eq::eos}, \eqref{eq::muj} and \eqref{eq::njTj}
\bsub\label{eq::bgevol}
\begin{eqnarray}
  && \dot{\mu}_{(i)} = -\Theta(\mu_{(i)} + p_{(i)})\ , \\\label{eq::Tdot}
  &&\dot{T}_{(i)}=-\tfrac{2}{3}\Theta T_{(i)}\ ,\\\label{eq::pdot}
  && \dot{p}_{(i)} = -\tfrac{5}{3}\Theta p_{(i)}\ . 
\end{eqnarray} 
\esub

The remaining variables are 
gauge invariant covariant ({\sc gic}) first order perturbations on the {\sc frw} background by the lemma in \cite{stewartlemma}. 
For an extensive physical motivation of using gauge invariant variables that vanish in the (fictitious) background see \citet{bardeen, ellis-bruni, Ellis-vanElst}. 

In the next section we will define {\sc gic} variables for all the physical quantities that govern the two fluids separately, and their mutual interactions and derive covariant evolution equations for all these variables.

%-----------------------------------------
\section{Evolution of GIC vector perturbations}\label{sec::vector}
We now proceed to find evolution equations for the combined two-fluid as determined by the observers $u^{a}$ (note that until now this frame is still left arbitrary). 
To facilitate this we define the variables in Table~\ref{table::sumdif}.

\begin{table}
\caption{Variables for the summed and subtracted quantities for the two fluid species.}
\label{table::sumdif}\centerline{
\begin{tabular}{l|cc}\hline
&Total  & Differential\\\hline
&&\\[-.5em]
Energy dens.&$\mu = \mu_{(1)} + \mu_{(2)}$&$ \delta\mu = \mu_{(1)} - \mu_{(2)}$\\[.5em]
Pressure &$p = p_{(1)} + p_{(2)}$&$ \delta p = p_{(1)} - p _{(2)}$\\[.5em]
Enthalpy &$h = h_{(1)} + h_{(2)}$&$ \delta h = h_{(1)} - h _{(2)}$\\[.5em]
Temperature &$T = T_{(1)} + T_{(2)}$& $\delta T = T_{(1)} - T _{(2)}$\\[.5em]
Number dens.&$N = N_{(1)} + N_{(2)}$& $\delta n = n_{(1)} - n _{(2)}$\\[.5em]
Velocity &$V = \tfrac{1}{2}(v_{(1)} + v_{(2)})$&$ \delta v = \tfrac{1}{2}(v_{(1)} - v _{(2)})$\\[.5em]
Mass &$M = m_{(1)} + m_{(2)}$&$ \delta m= m_{(1)} - m _{(2)}$\\[.5em]\hline
\end{tabular}}
%\begin{align}\nonumber
%\mu &= \mu_{(1)} + \mu_{(2)}& \delta\mu &= \mu_{(1)} - \mu_{(2)},\\\nonumber
%p &= p_{(1)} + p_{(2)}& \delta p &= p_{(1)} - p _{(2)}\\\nonumber
%h &= h_{(1)} + h_{(2)}& \delta h &= h_{(1)} - h _{(2)}\\\nonumber
%T &= T_{(1)} + T_{(2)}& \delta T &= T_{(1)} - T _{(2)}\\\nonumber
%N &= N_{(1)} + N_{(2)}& \delta n &= n_{(1)} - n _{(2)}\\\nonumber
%V &= \tfrac{1}{2}(v_{(1)} + v_{(2)})& \delta v &= \tfrac{1}{2}(v_{(1)} - v _{(2)})\\\nonumber
%M &= m_{(1)} + m_{(2)}& \delta m &= m_{(1)} - m _{(2)}
%\end{align}
\end{table}
In terms of the variables in Table~\ref{table::sumdif} we can write the first order charge and current densities as:
\begin{align}\label{eq::chargecurrent}
\rho &= e\delta n\ ,& 
 j^a &= e N\delta v^a\ .
 \end{align}
Or alternatively, the velocity difference can be interpreted as a normalised current density $\delta v^{a} = j^{a}/(e N)$ and similarly, the number density difference $\delta n$ is clearly equivalent to the charge density $\delta n = \rho/e$.

The evolution equations for the density and temperature variables (with $\delta T\neq 0 $ in the background) follow from \eqref{eq::njTj} and \eqref{eq::Tdot}:
\begin{subequations}\label{eq:Ndndot}
\begin{eqnarray}\label{eq:dndot}
  && \frac{\dot{\delta n}}{N} = -\Theta \frac{\delta n}{N} - \D_a\delta v^a , \\\label{eq:Ndot}
  && \frac{\dot{N}}{N} = - \Theta - \D_a V^a , \label{eq:numdensity}\\
  && \frac{\dot{\delta T}}{T} = -\tfrac{2}{3}(\Theta+\D_aV^a)\frac{\delta T}{T}  -\tfrac{2}{3}\D_a\delta v^a\ , \\
  && \frac{\dot{T}}{T} = -\tfrac{2}{3}(\Theta  + \D_aV^a)  -\tfrac{2}{3}(\D_a\delta v^a)\frac{\delta T}{T}\ . 
  \label{eq:tottemp} 
\end{eqnarray}
\end{subequations}

Similarly, we find the total momentum conservation equation by summing \eqref{eq::vj} over the two species
\begin{equation}\label{eq::acc}
  h\dot{u}^a = -\D^ap - \tfrac{4}{3}\Theta q^a - \dot{q}^{\langle a\rangle}\ ,
\end{equation}
while subtracting \eqref{eq::vj} yields
\begin{align}\nonumber
  \delta h \dot{V}^{\langle a\rangle}
  + h \dot{\delta v}^{\langle a\rangle} 
  &= - (\tfrac{\Theta}{3} \delta h + \delta\dot{ p})V^a 
    - (\tfrac{\Theta}{3} h+\dot{p})\delta v^a 
   \\\label{eq::curr}
    & - \delta h\ \dot{u}^a  - \D^a\delta p -\delta\rho E^a\ .
\end{align}

\subsection{Frame choice}
From hereon it will be convenient to specify to a particular observer frame $u^{a}$. Customary choices are either the {\em particle frame} in which to linear order $\sum n_{(i)} v_{(i)} = N V^{a} =0$ such that the total velocity perturbation with respect to the observer always vanishes, or the {\em energy frame} in which $q^{a}= \sum h_{(i)} v_{(i)} = h V^{a} + \delta h \delta v^{a} =0$. 

Both are useful in eliminating either $V^{a}$ or $\delta v^{a}$. In the particle frame, this relation is particularly simple with $V^{a} = 0$ and $\delta v = v_{(1)} = v_{(2)}$. However, requiring the total velocity perturbation to vanish when studying wavelength dependent perturbations in the total density, for instance, doesn't seem like an obvious frame choice. Choosing the energy frame means that the acceleration is driven only by pressure gradients: $h \dot{u}^{a} = - \D^{a}p$ from the total momentum equation \eqref{eq::acc}. The relation $V^{a} = \delta v^{a}(\delta h/h)$ is convenient in a cold plasma when $h_{(i)} = m_{(i)} n_{(i)}$ and $\delta h/h$ is time independent. In our case, however, when the pressure contributes to the energy density, this coefficient has a complicated temporal dependence and doesn't really simplify the calculations.

In this paper we therefore prefer to choose the {\em geodesic frame} of freely-falling observers such that $\dot{u}^{a}=0$ (and $q^{a}\neq 0$). The corresponding $4$-velocity is covariantly defined and reduces to the perfect fluid flow velocity of the fundamental observers in the background. Therefore the {\sc gi} variables defined by gradients orthogonal to the fluid flow are properly defined and the geodesic frame is a valid choice.

One could interpret \eqref{eq::acc} as defining $\delta v^{a}$ as a function of $V^{a}$ and eliminate $V^{a}$ in favour of $\delta v^{a}$ in \eqref{eq::curr} to get an equation similar to those in the energy frame. 
Rather, we treat \eqref{eq::acc} and \eqref{eq::curr} on equal footing as two coupled differential equation for the total velocity and the normalised current density.
By eliminating $\dot{p}$ and $\delta\dot{p}$ using \eqref{eq::pdot}, we find from \eqref{eq::acc} and \eqref{eq::curr} the (almost) symmetrical relations
\bsub\label{eq::vV}
\begin{align}\nonumber
  h \dot{V}^{\langle a\rangle}
  +  \delta h \dot{\delta v}^{\langle a\rangle} 
  &= - \tfrac{\Theta}{3}(  \mu - 4  p)V^a 
    - \tfrac{\Theta}{3}(  \delta\mu-4 \delta p)\delta v^a \\\label{eq::v1}
    &- \D^a p\ .\\\nonumber
  \delta h \dot{V}^{\langle a\rangle}
  + h \dot{\delta v}^{\langle a\rangle} 
  &= - \tfrac{\Theta}{3}( \delta \mu - 4 \delta p)V^a 
    - \tfrac{\Theta}{3}( \mu-4p)\delta v^a 
   \\\label{eq::v2}
    & -    \D^a\delta p -\delta\rho E^a\ .
\end{align}
\esub
The above equations are clearly {\sc gic} since $V^{a}, \delta v^{a}, \D^{a}p, \D^{a}\delta p$ and $E^{a}$ all vanish in the background (implying that all the coefficients in \eqref{eq::vV} should be evaluated in the background).

\subsection{Definition of GIC vector quantities}\label{sec::gicphysical}
It is argued in \cite{ellis-bruni} that the physically relevant variables characterising the spatial variation of scalar quantities that do not vanish in the background (such as the energy density) are the {\em dimensionless comoving fractional spatial gradients} of those quantities. In particular, we define {\sc gic} variables for the density, temperature, expansion and velocity perturbations in Table~\ref{table::gicvar}. 
\begin{table}
\caption{{\sc gic} dimensionless comoving scalar variables.}
\label{table::gicvar}\centerline{
\begin{tabular}{cc}\hline
Total & Differential \\\hline
&\\[-.5em]
$X^{a}= S\frac{\D^{a}N}{N}$&$ x^{a}= S\frac{\D^{a}\delta n}{N}$\\[.5em]
$Y^{a}= S\frac{\D^{a}T}{T}$&$ y^{a}= S\frac{\D^{a}\delta T}{T}$\\[.5em]
$W^{a}= S\D^{a}\D_{b}V^{b}$&$ w^{a}= S\D^{a}\D_{b} \delta v^{b}$\\[.5em]
$Z^{a}= S\D^{a}\Theta$ \\[.5em]\hline
\end{tabular}}
%\begin{align}\nonumber
%X^{a}&= S\frac{\D^{a}N}{N}& x^{a}&= S\frac{\D^{a}\delta n}{N}\\\nonumber
%Y^{a}&= S\frac{\D^{a}T}{T}& y^{a}&= S\frac{\D^{a}\delta T}{T}\\\nonumber
%W^{a}&= S\D^{a}\D_{b}V^{b}& w^{a}&= S\D^{a}\D_{b} \delta v^{b}\\\nonumber
%Z^{a}&= S\D^{a}\Theta&&
%\end{align}
\end{table}

The variables such as $X^{a}$ are the quantities that a typical observer would measure. For instance, at a particular time one can measure the comoving gradient in the number density (with respect to the length-scale given by the background scale-factor $S$) by simply counting the number of distant galaxies or clusters of galaxies at each distance (obviously, closer by sources like those in our own galaxy do not follow the `cosmic fluid' and are neither homogeneous nor isotropic). Other choices like $\delta\mu/\mu$ comparing the perturbed energy density with its background value depend on the fictitious background and are gauge dependent to the extent that some perturbations can be eliminated by a choice of gauge and the distinction between physical modes and spurious gauge modes can be obscured.

The velocity variables such as $\D^{a}(\D_{b}V^{b})$ might be interpreted as the gradient of some kind of fluid compression, but in this paper will be treated simply as auxiliary variables needed to couple \eqref{eq::vV} to \eqref{eq:Ndndot}.

\subsection{Evolution of GIC vector perturbations}

To obtain {\sc gic} evolution equations for vector perturbations in the total and relative densities and temperatures and the expansion we take the comoving gradients of \eqref{eq:Ndndot} which conveniently commute with the time derivative (this follows from the commutation relations in the Appendix \eqref{eq::commutations} and the definition \eqref{eq::scale}). The expansion terms proportional to $\Theta$ in \eqref{eq:Ndndot} are absorbed by choosing dimensionless variables. For example from \eqref{eq:dndot} and \eqref{eq:Ndot} we find (to linear order):
\[
\left(\frac{\delta n}{N}\right)^{\cdot}
=\frac{\delta \dot{n}}{N} + \Theta \frac{\delta n}{N}\ .\]

The resulting {\sc gic} evolution equations for the vector quantities in Table~\ref{table::gicvar} are
\bsub\label{eq::gicvector}
\bqa
\dot{X}^{\langle a \rangle}&=& - [W^{a}+ Z^{a}]\ ,\\\label{eq::xdot}
\dot{x}^{\langle a \rangle}&=& - w^{a}\ ,\\
\dot{Y}^{\langle a \rangle}&=& - \tfrac{2}{3}[W^{a}+\alpha w^{a}+Z^{a}]\ ,\\
\dot{y}^{\langle a \rangle}&=& - \tfrac{2}{3}[\alpha W^{a}+ w^{a}+\alpha Z^{a}]\ ,\\\nonumber
\dot{Z}^{\langle a \rangle}&=&- \tfrac{2}{3}\Theta Z^{a}- \tfrac{\kappa M N}{4} 
\left[\tfrac{9}{2} \bar{T} Y^{a} \right. \\ 
&+&\left.\left(1+ \tfrac{9}{2} \bar{T}\right) X^{a}+ 
\left(\bar{m}+ \tfrac{9}{2}\alpha \bar{T}\right) x^{a} \right]\ ,
\eqa
\esub
where $\bar{m} = \delta m/M$, $\bar{T} = T/M$ and $\alpha = \delta T/T$. Note that \eqref{eq::gicvector} are still valid in any observer frame $u^{a}$ and also that $\alpha$ is constant in the background.

We define $\Omega^{a}=\curl V^{a}$, $\omega^{a}=\curl \delta v^{a}$, and ${\cal C}^a = \curl E^a$ and take the curl of \eqref{eq::vV}. The pressure gradients vanish and using \eqref{eq::commutations} we find
\begin{align}
&h \dot{\Omega}^{a} + \delta h \dot{\omega}^{a} =
\Theta\left[(p-\tfrac{2}{3}\mu)\Omega^{a}+(\delta p-\tfrac{2}{3}\delta\mu)\omega^{a} \right]\ ,\\\nonumber
&h \dot{\omega}^{a} + \delta h \dot{\Omega}^{a} =
\Theta\left[(p-\tfrac{2}{3}\mu)\omega^{a}+(\delta p-\tfrac{2}{3}\delta\mu)\Omega^{a} \right]
+ e N {\cal C}^{a}\ .\hfill
\end{align}

The curl-free part of the velocity perturbations follows from taking the divergence of \eqref{eq::vV} and eliminating $\D_{a}E^{a}$ in favour of a charge density perturbation and the background plasma frequency
\begin{align}\label{eq::omegap}
\omega_{p}^{2}= \sum_{(i)}\omega_{p,(i)}^{2} = \frac{e^{2}}{\epsilon_{0}}\left(\frac{n_{(1)}}{m_{(1)}}+\frac{n_{(2)}}{m_{(2)}} \right) = \frac{e^{2}N}{\epsilon_{0}M}\frac{2}{1-\bar{m}^{2}}\ .&
\end{align}
More important though is the gradient of the resulting equations, which yields the equations for $W^{a}$ and $w^{a}$ that couple back to the density and temperature perturbations \eqref{eq::gicvector}.
These {\sc gic} auxiliary velocity variables evolve as:
\bsub
\begin{align}\nonumber
&h \dot{W}^{\langle a \rangle} + \delta h \dot{w}^{\langle a \rangle}=\\\label{eq::W}
&-\tfrac{N T}{2}  \left[\D^{2} + \tfrac{2}{9}(\Theta^{2}-3\mu) \right]
[X^{a}+Y^{a}+\alpha x^{a}] \\\nonumber
&- \Theta 
\left[\left(\tfrac{2}{3}\mu -p \right)W^{a}+
\left(\tfrac{2}{3}\delta\mu -\delta p \right)w^{a}\right] \ ,
\end{align}and
\begin{align}\nonumber
&h \dot{w}^{\langle a \rangle} + \delta h \dot{W}^{\langle a \rangle}= (1-\bar{m}^{2})\tfrac{N M\omega_{p}^{2}}{2} x^{a}\\\nonumber
&-\frac{N T}{2}  \left[\D^{2} + \tfrac{2}{9}(\Theta^{2}-3\mu) \right]
[\alpha X^{a}+y^{a}+ x^{a}] \\\label{eq::w}
&- \Theta 
\left[\left(\tfrac{2}{3}\mu -p \right)w^{a}+
\left(\tfrac{2}{3}\delta\mu -\delta p \right)W^{a}\right]\ ,
\end{align}
\esub
where we have used \eqref{eq:divE}, the commutation relations \eqref{eq::commutations} and
\bsub
\bqa
\D^{a}p &=& \frac{N T}{2} (X^{a}+ \alpha x^{a} +Y^{a})\ ,\\
\D^{a}\delta p &=& \frac{N T}{2} (\alpha X^{a}+ x^{a}+y^{a})\ .
\eqa
\esub

\subsection{GI electromagnetic equations}
To investigate how the velocity perturbations couple to the electromagnetic field we need the solenoidal part of \eqref{eq::v2}, which couples to $\dot{B}^{a}$ through \eqref{eq::dotB} and the curl-free scalar part which through \eqref{eq:divE} couples to the plasma frequency \eqref{eq::omegap}.

From the constraint \eqref{eq:divE} and \eqref{eq::xdot} we trivially find an evolution equation related to the curl free part of the electric field in terms of 
$S \D^a (\D_b E^b) \equiv \mathcal{E}^a = e N x^{a}/\epsilon_{0}$
\bq
  \dot{\mathcal{E}}^{\langle a \rangle} = - \frac{eN}{\epsilon_0}w^a\ .
\eq
From Maxwell's equations, we find (in agreement with \cite{Tsagas}) the wave equations to linear order 
\bsub
\bqa\nonumber
  \ddot{E}^{\langle a\rangle} - \D^2E^a &=& 
  \tfrac{1}{3}(\mu + 3p)E^a - \tfrac{5}{3}\Theta\dot{E}^{\langle a\rangle}
  - \tfrac{4}{9}\Theta^2E^a \\\label{eq:Ewave}
  &&- \frac{e}{\epsilon_0}\D^a\delta n 
  - \mu_0eN\dot{\delta v}^{\langle a\rangle}\ ,
\\\nonumber
  \ddot{B}^{\langle a\rangle} - \D^2B^a &=& 
  \tfrac{1}{3}(\mu + 3p)B^a - \tfrac{5}{3}\Theta\dot{B}^{\langle a\rangle}
  - \tfrac{4}{9}\Theta^2B^a \\\label{eq:Bwave}
  &&+ \mu_0eN\omega^a\ .
\eqa
\esub
The expansion normalised curl of the electric field $\mathcal{C}^a = S\ \curl\,E^a$ satisfies the wave equation
\begin{align}\nonumber
   \ddot{\mathcal{C}}^{\langle a\rangle} - \D^2\mathcal{C}^a + \tfrac{5}{3}\Theta\dot{\mathcal{C}}^{\langle a\rangle}
&= 
  \tfrac{1}{3}(\mu + 3p - \tfrac{4}{3}\Theta^{2})\mathcal{C}^a  \\  
  &- \mu_0eN\left[\dot{\omega}^{\langle a\rangle} + \tfrac{\Theta}{3}\omega^a\right],
\end{align}
from (\ref{eq:Ewave}) and (\ref{eq::commutations}).

The evolution equation derived in this section together with the constraints completely determine the behaviour of {\sc gic} vector perturbations of the two-temperature plasma. 
Here we will focus on scalar perturbations. In the next section we will derive the corresponding evolution equations.

%-----------------------------------------
\section{GIC scalar perturbations}\label{sec::scalar}
The easiest way to obtain a closed set of differential equations for scalar perturbations of the two-temperature plasma is to take the divergence of the vector equations derived in the previous section. 
The resulting {\sc gic} expansion normalised dimensionless variables are summarised in Table~\ref{table::scalars}.

In section~\ref{sec::gicphysical} we argued that the variables in Table~\ref{table::gicvar}, which are dimensionless fractional spatial gradients of scalars quantities that do not vanish in the background space-time, are in fact the physically relevant variables that an observer would measure directly.  By taking another divergence of these variables and subsequently performing a harmonic decomposition (in the next section), we essentially introduce a length-scale $\nabla^{2} \sim L^{-2}$ corresponding to a Fourier component of a perturbation with wavenumber $k = 2\pi/L$. Thus $\Delta$, for instance,  represents a comoving density perturbation on a length-scale $L$:    
$\Delta \sim (S/L)^{2} (\delta \rho/\rho)$, which is similar to 
the variable $\delta \rho/\rho$ that is sometimes used, but defined to be covariant and gauge invariant. A similar interpretation applies to the other variables in Table~\ref{table::scalars}.

\begin{table}
\caption{Gauge invariant scalar expansion normalized variables.}
\label{table::scalars}\centerline{
\begin{tabular}{cc}
\hline
Total & Differential \\\hline
&\\[-.5em]
$\Delta = S^{2}\frac{\D^{2}N}{N}$&$ \delta= S^{2}\frac{\D^{2}\delta n}{N}$\\[.5em]
$\Gamma = S^{2}\frac{\D^{2}T}{T}$&$ \gamma = S^{2}\frac{\D^{2}\delta T}{T}$\\[.5em]
${\cal W} = S^{2}\D^{2}\D_{b}V^{b}$&$ {\cal V} = S^{2}\D^{2}\D_{b} \delta v^{b}$\\[.5em]
$\zeta = S^{2} \D^{2}\Theta$&\\[.5em]\hline
\end{tabular}}
\end{table}

Since the comoving divergence again commutes with the time derivative, the scalar equations look very similar to \eqref{eq::gicvector}. All the coefficients have to be zeroth order and it is more illustrative to re-write those in terms of the background quantities $N$, $T$, $\rho_{m} = M N/2$ etc. The resulting evolution equations are
\bsub\label{eq::first}
\bqa\label{eq::Ddot}
\dot{\Delta}&=& - [{\cal W} + \zeta ]\ ,\\\label{eq::ddot}
\dot{\delta}&=& - {\cal V}\ ,\\\label{eq::Gdot}
\dot{\Gamma}&=& - \tfrac{2}{3}[{\cal W}+\alpha {\cal V}+\zeta]= \tfrac{2}{3}[\alpha\dot{\delta} + \dot{\Delta}]\ ,\\\label{eq::gdot}
\dot{\gamma}&=& - \tfrac{2}{3}[\alpha {\cal W} + {\cal V} +\alpha \zeta] = \tfrac{2}{3}[\alpha\dot{\Delta} + \dot{\delta}]\ ,\\\nonumber 
\dot{\zeta}&=&- \tfrac{2}{3}\Theta \zeta - \tfrac{1}{2}\kappa \rho_{m}\left[\tfrac{9}{2} \bar{T} \Gamma \right. \\\label{eq::zdot}
&+&\left.\left(1+ \tfrac{9}{2} \bar{T}\right) \Delta + 
\left(\bar{m}+ \tfrac{9}{2}\alpha \bar{T}\right) \delta\right]\ .
\eqa
\esub
In comparing with the non-thermal dust results in \cite{betschart04} we find that (\ref{eq::ddot}, \ref{eq::Ddot}) agree with their (24a,d). For the latter identification one should transform back from the energy frame where $V^{a} = - \delta v^{a} \delta\mu/\mu$ was used to eliminate $V^{a}$, and instead leave the equation in terms of $\delta v^{a}$. Clearly the equations for the thermal perturbations are new and the evolution equations for the velocity and expansion perturbations are more complicated when including temperature effects.

\subsection{Harmonic decomposition}
In terms of the variables in Table~\ref{table::scalars}, the evolution of the scalar velocity variables is found by taking the divergence of (\ref{eq::W}, \ref{eq::w}) using \eqref{eq::commutations}.\footnote{The commutation rule for $\D_{a}\D^{2}$ cancels the factor proportional to the expansion in the previous $\D^{a}\D^{2}$ commutator.} Furthermore, we harmonically decompose the Laplacians in terms of the scalar harmonic functions $Q^{(\ell)}$ that satisfy the Helmholtz equation
\bq
\D^{2}Q^{(\ell)}=- \frac{\ell^{2}}{S^{2}}Q^{(\ell)}\ ,\quad
\dot{Q^{(\ell)}} = 0\ ,\quad f  = \sum_{\ell} f_{(\ell)} Q^{(\ell)}\ ,
\eq
where $f$ is any perturbation scalar and $f_{(\ell)}$ the corresponding harmonic \citep{harrison67}. For the sake of clarity we will drop the index $\ell$, as we will not encounter harmonic mixing due to the linearity of the equations. Each evolution equation is understood to represent the temporal behaviour in one harmonic mode $\ell$. 
The {\em comoving wavelength} of a perturbation is given in terms of the wave-number $\ell$ by $\lambda(\tau) = 2\pi S(t)/\ell$ where time dependencies are given in terms of $\tau = t/t_{0}$, the proper time along a world-line normalised to some initial time $t_{0}$. The velocity variables then evolve as
\bsub
\begin{align}\label{eq::SS}
&\left(1+\tfrac{5}{2}\bar{T}\right)\dot{{\cal W}} + \left(\bar{m}+\tfrac{5}{2}\alpha \bar{T}\right)\dot{{\cal V}} =\\\nonumber
&\qquad\frac{\ell^{2}\bar{T}}{ S^{2}}( \Delta + \alpha \delta+\Gamma) - \tfrac{2}{3} \Theta({\cal W} + \bar{m}{\cal V})\ ,\\\label{eq::ss}
&\left(1+\tfrac{5}{2}\bar{T}\right)\dot{{\cal V}} + \left(\bar{m}+\tfrac{5}{2}\alpha \bar{T}\right)\dot{{\cal W}} =\\\nonumber
&\qquad\frac{\ell^{2}\bar{T}}{ S^{2}}(\alpha \Delta + \delta+\gamma)
- \tfrac{2\Theta}{3} ({\cal V} + \bar{m}{\cal W})
+(1-\bar{m}^{2})\omega_{p}^{2}\delta\ .
\end{align}
\esub

\subsection{Sound velocities}\label{sec::sound}
We assume that the background plasma is non-relativistic such that the total thermal energy is small compared to the total rest-mass energy. Consequently, $\bar{T} = T/M \ll 1$ and we will neglect $\bar{T}$ in the gravitational interactions 
and use $\mu \simeq \mu + p\simeq \rho_{m}$ in the background.

We define the sound velocity for each species as:
\bq
c_{s,(i)}^{2}= \frac{\dot{p}_{(i)}}{\dot{\mu}_{(i)}} \simeq \frac{\dot{n}_{(i)}T_{(i)} + n_{(i)}\dot{T}_{(i)}}{m_{(i)} \dot{n}_{(i)}} = \frac{5}{3}\frac{T_{(i)}}{m_{(i)}}\ ,
\eq
using \eqref{eq::bgevol}. We define total and differential sound speeds similar to $V^{a}$ and $\delta v^{a}$ as
\bsub\label{eq::soundspeeds}
\bqa\label{eq::soundtotmy}
C_{s}^{2} &=& \tfrac{1}{2} (c_{s,1}^{2} + c_{s,2}^{2})=\tfrac{5}{3}\bar{T}\left( \frac{1-\bar{m}\alpha}{1-\bar{m}^{2}}\right),\\\label{eq::dcs}
\delta c_{s}^{2} &=&\tfrac{1}{2} (c_{s,1}^{2} - c_{s,2}^{2})= \pm \tfrac{5}{3}\bar{T} \left(\frac{\alpha-\bar{m}}{1-\bar{m}^{2}}\right),
\eqa
\esub
where the sign of \eqref{eq::dcs} is positive when $c_{s,1}^{2} > c_{s,2}^{2}$ and negative when $c_{s,1}^{2} < c_{s,2}^{2}$. The variable $C_{s}^{2}$ is related to the {\em physical} mass-weighted total sound velocity $c_{s}^{2} = \sum_{(i)} c_{(i)}^{2} \dot{\mu}_{(i)}/\dot{\mu} = 5 p/3\mu$ (for our particular equation of state) by
\bq\label{eq::realsound}
C_{s}^{2} = c_{s}^{2} \left( \frac{1-(\delta\mu/\mu)(\delta p /p)}{1-(\delta\mu/\mu)^{2}}\right) = c_{s}^{2} \left( \frac{1-\bar{m}\alpha}{1-\bar{m}^{2}}\right)\ ,
\eq
where the second expression ($c_{s}^{2} = 5\bar{T}/3$) follows from the fact that  the number densities are the same in the background. In the remainder of this paper it will be apparent that $C_{s}^{2}$ in \eqref{eq::soundtotmy} is a more convenient variable than $c_{s}^{2}$ in \eqref{eq::realsound} because the former only depends only on the temperature whereas the latter depends on both the temperature (pressure) and number density. The results will only depend on the (physical) sound speed of one of the two species.

\subsection{Wave equations}
In the non-relativistic limit and using the definitions in the previous section, the evolution equations for the expansion and the auxiliary variables ${\cal W}$ and ${\cal V}$ are given by
\bsub\label{eq::evol}
\begin{align}\label{eq::zevol}
&\dot{\zeta}+ \tfrac{2}{3}\Theta \zeta =- \tfrac{1}{2}\kappa \rho_{m}\left[\Delta + \bar{m} \delta\right]\ ,\\\label{eq::Sevol}
&\dot{{\cal W}} + \tfrac{2}{3}\Theta {\cal W} = \frac{3\ell^{2}C_{s}^{2}}{5S^{2}}\left[\Delta +\frac{\delta c_{s}^{2}}{C_{s}^{2}} \delta+\frac{\Gamma - \bar{m}\gamma}{1-\bar{m}\alpha} \right]- \bar{m} \omega_{p}^{2}\delta \ , \\\label{eq::sevol}
&\dot{{\cal V}}+\tfrac{2}{3}\Theta {\cal V}=\frac{3\ell^{2}C_{s}^{2}}{5S^{2}}\left[\delta +\frac{\delta c_{s}^{2}}{C_{s}^{2}} \Delta+\frac{\gamma - \bar{m}\Gamma}{1-\bar{m}\alpha} \right]+ \omega_{p}^{2}\delta  \ ,
\end{align}
\esub
where in this limit the expansion evolution \eqref{eq::zevol} does reduce to equation (24b) in \cite{betschart04} for the gravitational attraction of pressure-less dust.

Since we want to study wave-like harmonic perturbations of the density and temperature we take the second time derivative of (\ref{eq::Ddot}--\ref{eq::gdot}) and use the evolution equations \eqref{eq::evol} to eliminate $\zeta$, ${\cal V}$ and ${\cal W}$ from the second order differential equations for $\delta$, $\Delta$, $\gamma$ and $\Gamma$
\bsub\label{eq::second}
\bqa\label{eq::Dddot}
\ddot{\Delta}(\tau)&=& - [\dot{{\cal W}}(\tau) + \dot{\zeta}(\tau) ]\ ,\\\label{eq::dddot}
\ddot{\delta}(\tau)&=& - \dot{{\cal V}}(\tau)\ ,\\\label{eq::Gddot}
\ddot{\Gamma}(\tau)&=& - \tfrac{2}{3}[\dot{{\cal W}}(\tau)+\alpha \dot{{\cal V}}(\tau)+\dot{\zeta}(\tau)]\ ,\\\label{eq::gddot}
\ddot{\gamma}(\tau)&=& - \tfrac{2}{3}[\alpha \dot{{\cal W}}(\tau) + \dot{{\cal V}} (\tau)+\alpha \dot{\zeta}(\tau)]\ .
\eqa
\esub

Equivalently, we can simultaneously solve the combined set of first order ordinary differential equations \eqref{eq::first} and \eqref{eq::evol}. The solutions for initial conditions and limiting cases of interest are studied in the next section.

%-----------------------------------------
\section{Solutions}
In studying inhomogeneities as perturbations on a {\sc frw} Universe filled with an unmagnetized cold dusty plasma we expect the perturbations to behave as eigenmodes of such a plasma altered by the background curvature. 
We will therefore study the longitudinal Langmuir mode,
 ion-acoustic oscillations and the gravitational Jeans instability.

\subsection{General discussion}
By examining \eqref{eq::evol} we can make the following  observations:
\begin{itemize}
\item From \eqref{eq::zevol} we find that an initial perturbation in either the total density {\em or} the charge density causes a (gravitational) perturbation of the expansion, unless the plasma consists of electrons and positrons with $\bar{m}=0$ (in which case only a total density perturbation couples to the expansion);
\item A perturbation in the charge density corresponds to a longitudinal electric field perturbation (such that $\D_{a}E^{a}\neq 0$) which enters the equations of motion as a Lorentz force term and excites perturbations in the current density and (unless the plasma is $e^{\pm}$) the total velocity variables. This is expressed in (\ref{eq::Sevol}--\ref{eq::sevol}) in terms of the plasma frequency;
\item When the electric field perturbation is unimportant, only the wavelength dependent pressure gradient terms contribute in (\ref{eq::Sevol}--\ref{eq::sevol}). When the sound velocities of the two species are equal in the background, a perturbation in the total density only drives a perturbation in the total velocity, while a perturbation in the charge density only excites a current density perturbation. If the plasma also were $e^{\pm}\Rightarrow \bar{m}=0$ then (\ref{eq::Sevol}--\ref{eq::sevol}) completely de-couple and we have ${\cal W} = {\cal W}(\Delta, \Gamma)$ and ${\cal V} = {\cal V}(\delta, \gamma)$. However, in the general case that we want to investigate in this paper where the sound velocities in the background are allowed to differ, a perturbation in the charge density will excite a total velocity perturbation and more interestingly {\em a total density perturbation will excite a current density} from \eqref{eq::sevol}. 
\end{itemize}
The excited modes in $\zeta$, ${\cal W}$ and ${\cal V}$ derived from \eqref{eq::evol} feed into \eqref{eq::second}, or equivalently \eqref{eq::first} and we can make some remarks on how they drive perturbations, particularly in the temperature variables. 
\begin{itemize}
\item The behaviour of the temperature variables can be found most readily by integrating the second expressions in (\ref{eq::Gdot}--\ref{eq::gdot}) and setting the integration constants to match the appropriate initial conditions. For a one-temperature background we find that the total temperature is simply proportional to the total density, whereas a temperature difference only couples to a perturbation in the charge density.
\item It again is clear from (\ref{eq::Gddot}--\ref{eq::gddot}) that a temperature difference in the background allows for additional coupling between the equations. In particular, a perturbation in the total velocity or the expansion excites perturbations in the temperature difference. For instance, a growing mode in the expansion such as gravitational collapse can lead to a {\em growing temperature difference}.
\end{itemize}
This concludes our general discussion of the allowed excitations and couplings of the scalar perturbation variables in a non-trivial background. We now proceed to study some analytical solutions of these differential equations in more detail.

\subsection{Evolution of the background}
In solving the evolution equations for the perturbations we also have to include the evolution of the coefficients in the background. In the approximation that we introduced in Section~\ref{sec::sound} where the pressure is included in the sound velocity but not in the gravitational mass we can just use the well known evolution of a dust dominated Universe. 
The scale factor then evolves as $S(\tau)\propto \tau^{2/3}$ and the dimensionless expansion and gravitational mass as $\Theta = 2/\tau$ from \eqref{eq::scale} and $\kappa M N/4 = \kappa \mu/2 = 2/(3\tau^{2})$ from \eqref{eq::thetadot}. The number density, which is hidden in the plasma frequency, evolves as $N/N_{0} = \omega_{p}^{2} /\omega_{p0}^{2}=\tau^{-2}$ from (\ref{eq:Ndot}, \ref{eq::omegap}). 

Since the equation of state \eqref{eq::eos} is restricted to the non-relativistic regime we will not discuss the radiation dominated Universe with $p=\rho/3$ here, although the extension is straightforward and could be the subject of a future paper.

\subsection{Long wavelength Langmuir modes}\label{eq::longd}
The differential equations in the previous section are most readily solved in the long wavelength limit or alternatively the dust limit where $\bar{T} = 0$ everywhere. In both cases the terms proportional to $\bar{T}\ell^{2}/S^{2}$ vanish. 
The exact solutions derived in this section for $\Delta$ and $\delta$ agree with those in \cite{betschart04} (where $\delta$ is expressed in $Y=\delta n/N$) but we can now extend these results to include the possible thermal effects.

In this limit the wave equations for the charge density and total number density are
\bsub\label{eq::dD}
\begin{align}\label{eq::lngd}
&\ddot{\delta}(\tau) + \frac{4}{3 }\frac{\dot{\delta}(\tau)}{\tau}+ \omega_{p}^{2}\frac{\delta(\tau)}{\tau^{2}}=0\ ,\\\label{eq::lngD}
&\ddot{\Delta}(\tau) +  \frac{4}{3 }\frac{\dot{\Delta}(\tau)}{\tau}- \frac{2}{3} \frac{\Delta(\tau)}{\tau^{2}}=\bar{m}[\omega_{p}^{2}+ \tfrac{2}{3}] \frac{\delta(\tau)}{\tau^{2}}\ .
\end{align}
\esub
The first mode we consider is excited by an initial perturbation in the charge density $\delta(1) = \delta_{0}$ when all the other perturbations are zero initially. From \eqref{eq::first} we then find that
\[
\dot{\Delta}(1) = \dot{\delta}(1)=\dot{\Gamma}(1)=\dot{\gamma}(1)=0\ .
\]
The solution of \eqref{eq::lngd} subject to these initial conditions is
\bq\label{eq::langdsol}
\delta(\tau) =\delta_{0}\tau^{-\tfrac{1}{6}}\left[\cos(\omega\ln\tau) + \tfrac{1}{6\omega}\sin(\omega\ln\tau)\right]\ ,
\eq
where $\omega^{2} = \omega_{p}^{2}-1/36 \simeq  \omega_{p}^{2}$. This oscillatory solution agrees with \cite{betschart04}.
It is a non-propagating oscillation at the plasma frequency and is similar to the longitudinal Langmuir mode. The solutions are not simple plane waves but a more complicated harmonic oscillation with a decaying ($\tau^{-1/6}$) envelope due to the background expansion and corresponding evolution of the plasma frequency.

On inserting the solution \eqref{eq::lngd} in \eqref{eq::lngD} we can solve for $\Delta(\tau)$ to find
\bsub\label{eq::fulsol}
\bq\label{eq::langDsol}
\Delta(\tau) =  \bar{m}\delta_{0}\left(g(\tau) - \frac{\delta(\tau)}{\delta_{0}}\right)\ , 
\eq
with 
\bq\label{eq::grav}
g(\tau) = \frac{2}{5}\tau^{-1}+\frac{3}{5}\tau^{\frac{2}{3}}\ .
\eq
\esub
The solution \eqref{eq::fulsol} for the total density perturbation is a superposition of the oscillatory Langmuir mode and the usual growing and decaying (power-law) modes of the standard gravitational instability picture.
The growing mode always dominates (since $\tau >1$) leading to gravitational collapse on all length-scales. This is a well known result for dust since there is no pressure to act as a restoring force.

The temperature perturbations can be found by solving the wave equations (\ref{eq::Gddot}--\ref{eq::gddot}), but even simpler is to integrate the right-most expressions in (\ref{eq::Gdot}--\ref{eq::gdot}) taking care of the initial conditions, i.e.
\bsub\label{eq::langtemps}
\bqa\label{eq::langG}
\Gamma(\tau) &=& \tfrac{2}{3}\delta_{0}\left[(\alpha-\bar{m})\frac{\delta(\tau)}{\delta_{0}} - (\alpha-\bar{m} g(\tau))\right]\ ,\\\label{eq::dtgrows}
\gamma(\tau) &=&\tfrac{2}{3}\delta_{0}\left[(1-\bar{m}\alpha)\frac{\delta(\tau)}{\delta_{0}} - (1-\bar{m}\alpha g(\tau))\right]\ .
\eqa
\esub
From \eqref{eq::langtemps} we find that in an equal temperature plasma the relative temperature perturbation only fluctuates with the oscillation in the charge density, but by allowing for a temperature difference in the background ($|\alpha| > 0$) the {\em temperature difference grows as a power-law} during gravitational collapse. Note that from \eqref{eq::soundspeeds} $(\alpha-\bar{m}) \propto \delta c_{s}^{2}$, so the total temperature perturbation only couples to the charge density oscillation when the two fluids have unequal sound velocities in the background (by comparison $(1-\bar{m}\alpha) \propto  C_{s}^{2}$ in \eqref{eq::dtgrows}).

\subsection{Gravitationally driven temperature difference}\label{eq::longD}
We now study \eqref{eq::dD} in the same long wavelength or dust regime but start with the more common initial condition where the total density rather than the charge density is perturbed (so $\Delta(1) = \Delta_{0}$ and all the other perturbations vanish at $t=t_{0}$). With these initial conditions no perturbation of the charge density is excited and the total density only experiences the continuous gravitational collapse which couples to the temperature variables through the expansion \eqref{eq::evol}. The solutions are given in terms of the gravitational mode \eqref{eq::grav} by 
\bsub
\bqa
\delta(\tau)&=&0\ ,\\
\Delta (\tau) &=& \Delta_{0} g(\tau)\ ,\\\label{eq::gammajeans}
\gamma(\tau)&=& \tfrac{2}{3} \alpha \Delta_{0} (g(\tau)-1)\ ,\\
\Gamma(\tau)&=& \tfrac{2}{3}\Delta_{0} (g(\tau)-1)\ ,
\eqa
\esub
and we find from \eqref{eq::gammajeans} that even in this case {\em gravitational collapse increases an initial temperature difference} when we start with a perturbation in the total density such as those that are observed in the cosmic microwave background ({\sc cmb}).

\subsection{Wavelength dependent perturbations}
In treating perturbations of all wavelengths, the system of differential equations becomes substantially more complicated. We therefore specify to perturbations on time-scales that are small compared to the Hubble time such that the scale-factor is 
approximately constant $S(\tau) \sim S_{0}\Rightarrow \Theta = 0$ (or equivalently, the wavelengths are small with respect to the Hubble scale). Consequently, the $\kappa\rho_{m}/2$ terms vanish because of \eqref{eq::thetadot} and we can neglect the evolution of the coefficients ($N$, $\omega_{p}$ and $\bar{T}$) in the background. In this limit, the perturbations locally behave as plane waves and we can solve the system of equations algebraically and derive a dispersion relation.

For the wave equations we find in terms of $k_{0} = \ell/S(1)$
\bsub\label{eq::waves}
\begin{align}\label{eq::ddisp}
&\omega_{f}^{2}\delta =
\frac{3 k_{0}^{2}C_{s}^{2}}{5}\left[\delta +\frac{\delta c_{s}^{2}}{C_{s}^{2}} \Delta+\frac{\gamma - \bar{m}\Gamma}{1-\bar{m}\alpha} \right]+ \omega_{p}^{2} \delta\ ,\\
&\omega_{f}^{2}\Delta =
\frac{3 k_{0}^{2}C_{s}^{2}}{5}\left[\Delta +\frac{\delta c_{s}^{2}}{C_{s}^{2}} \delta+\frac{\Gamma - \bar{m}\gamma}{1-\bar{m}\alpha} \right]-\omega_{p}^{2} \bar{m} \delta\ ,
\end{align}
\esub
and from \eqref{eq::first} we simply have
\bsub
\begin{align}
&\Gamma = \tfrac{2}{3} \left[\alpha \delta + \Delta \right]\ ,\\
&\gamma = \tfrac{2}{3} \left[\alpha \Delta + \delta \right]\ .
\end{align}
\esub
A dispersion relation follows readily with solutions:
\begin{align}\nonumber
&\omega_{\pm}^{2} = k^{2}_{0} C_{s}^{2} + \tfrac{1}{2}\omega_{p}^{2}\left[1 \pm \sqrt{1 - \frac{4 k^{2}_{0} |\delta c_{s}^{2}|}{\omega_{p}^{2}}\left(\bar{m} - \frac{k^{2}_{0} |\delta c_{s}^{2}|}{\omega_{p}^{2}} \right)}\right]\ .
\end{align}
If the ion-acoustic frequency is small compared to the total plasma frequency, we can expand the square-root term in $k_{0}^{2}\delta c_{s}^{2}/ \omega_{p}^{2} \ll 1$ to find 
\bsub\bqa\label{eq::ionsound}
\omega_{-}^{2} &\simeq& k_{0}^{2} [C_{s}^{2} +  |\delta c_{s}^{2}|] = k_{0}^{2}\max(c_{s, 1},c_{s, 2})^{2} \ ,\\\nonumber
\omega_{+}^{2} &\simeq& k_{0}^{2} [C_{s}^{2} -  |\delta c_{s}^{2}|] + \omega_{p}^{2}\\\label{eq::langmuir}
&=& k_{0}^{2}\min(c_{s, 1},c_{s, 2})^{2} + \omega_{p}^{2}\ ,
\eqa\esub
which are the {\em ion-acoustic} and {\em Langmuir} modes, respectively.

\subsection{Jeans criterion}\label{sec::jeans}
To obtain a Jeans length-scale for collapse, the gravitational interaction is clearly essential. Therefore we cannot neglect $\kappa \rho_{m}$ which means we cannot neglect $\Theta$ and $\dot{\Theta}$ so we have to treat the full system of equations. The goal is to study the properties of the wave equation for the total density perturbations and find the critical length-scale distinguishing collapse from oscillatory behaviour. From the non-relativistic limit of the expansion equation \eqref{eq::zevol} and \eqref{eq::second} one realises that rather than to study perturbations in the total number density $\Delta$, a more physical choice is to study perturbations of the {\sc gi} normalised density variable
\bq
{\cal D} = \Delta  + \bar{m} \delta= S^{2}\frac{\D^{2} \rho_{m}}{\rho_{m}} \ ,
\eq
where $\rho_{m}= \tfrac{1}{2}(M N + \delta m \delta n)$ to first order. Furthermore, we have to eliminate the temperature variables $\Gamma$ and $\gamma$ from (\ref{eq::evol}, \ref{eq::second})
which is achieved by integrating \eqref{eq::first} to find
\bsub
\bqa
\Gamma - \bar{m}\gamma &=&\tfrac{2}{3}\left[(\alpha - \bar{m})\delta + (1-\bar{m}\alpha)\Delta \right]\ ,\\
\gamma - \bar{m}\Gamma &=&\tfrac{2}{3}\left[(\alpha - \bar{m})\Delta + (1-\bar{m}\alpha)\delta \right]\ .
\eqa
\esub
Putting all of this together, we find
\begin{align}\nonumber
&\ddot{{\cal D}} + \tfrac{2}{3}\Theta\dot{{\cal D}} + \left[\frac{\ell^{2}}{S^{2}}(C_{s}^{2} + \bar{m} |\delta c_{s}^{2}|) - \tfrac{1}{2}\kappa \rho_{m} \right] {\cal D}  =\\\label{eq::jeanss}
&\qquad\qquad \frac{\ell^{2}}{S^{2}} (1-\bar{m}^{2}) |\delta c_{s}^{2}| \delta\ .
\end{align}
We assume that the term on the right-hand-side of the equation is negligible for $\bar{m} \sim 1$. 
Note also that the plasma frequency dependent terms have dropped out in the equation for ${\cal D}$ which can be seen from (\ref{eq::evol}, \ref{eq::second}) or \eqref{eq::waves}.

Finding an exact solution of \eqref{eq::jeanss} is complicated by the fact that all the coefficients are themselves time dependent: $\Theta(\tau)$, $c_{s}(\tau)$, $\delta c_{s}(\tau)$, $S(\tau)$ and $\rho_{m}(\tau)$. However, a general result from differential calculus which in this particular case is called the {\em Jeans criterion} states that the solution for ${\cal D}$ changes from oscillatory to growing or decaying solutions when the term in square brackets changes sign (which corresponds to the perturbation frequency changing from real to imaginary).

 The limiting case is generally expressed as a Jeans wavelength $\lambda_{J}(\tau) = 2 \pi S(\tau)/\ell$ which corresponds to the length-scale at which the plasma becomes unstable to gravitational collapse. For length-scales smaller than $\lambda_{J}$ the pressure prevents collapse. 
For an electron-ion plasma with $\bar{m}\simeq 1$ we find
\bq\label{eq::jeanslength}
\lambda_{J}^{2} = \frac{\pi}{G \rho_{m}} (C_{s}^{2} + |\delta c_{s}^{2}|) \ ,
\eq
which corresponds to a well known result when $\delta c_{s}=0$ (see for instance equation (220) in \cite{Ellis-vanElst}). In our more general treatment we find, given \eqref{eq::soundspeeds}, that in a two-temperature plasma, the Jeans wavelength is determined by the species with the largest sound velocity. 
Note that, as in \cite{Ellis-vanElst} the Jeans length $\lambda_{J}$ is time-dependent. 

The corresponding {\em Jeans mass} is given by
\bq
M_{J} = \frac{4\pi \lambda_{J}^{3}}{3}\rho_{m} = \left[\frac{\pi}{G}\right]^{\tfrac{3}{2}}\frac{4 \pi \max(c_{s, 1},c_{s, 2})^{3}}{3}\sqrt{\rho_{m}} \ .
\eq

The limiting cases discussed in Sections~\ref{eq::longd}--\ref{eq::longD} are valid for wavelengths much longer that the Jeans wavelength \eqref{eq::jeanslength}.

%----------------------------------------%
\section{Conclusions}		%
%----------------------------------------%
In this paper we have extended the gauge invariant covariant perturbation theory on curved manifolds to include temperature effects in a multi-fluid that is not in complete thermal equilibrium. We have derived two closed sets of differential equations: one for gauge invariant covariant {\em vector} quantities in Section~\ref{sec::vector} and one for corresponding {\em scalar} variables in Section~\ref{sec::scalar}. The system in Section~\ref{sec::vector} could be the subject of further investigations, but in this paper we have focused on studying the behaviour of the scalar perturbations. 

In the long wavelength limit for a pressure-less plasma we recover the results in \cite{betschart04} but we have generalized the treatment to a two-fluid where both species are perfect fluids with an ideal gas equation of state and find that the total temperature and relative temperature perturbations follow the total and charge densities, respectively. If the two fluids furthermore have slightly different temperatures in the background we find that they also couple to each other. Most interestingly, the temperature difference grows during gravitational collapse. This result applies both to initial perturbations in the charge density or in the total density. We expect that the process of structure formation will be effected by this thermal evolution.

To find propagating oscillatory modes we have included the wavelength dependent behaviour but neglected the evolution of the background which allows us to treat the perturbations as locally plane waves. In this approximation we have algebraically obtained a dispersion relation with eigen-frequency solutions corresponding to Langmuir and ion-acoustic modes.

Finally,  in Section~\ref{sec::jeans} we have derived a modified Jeans wavelength that includes thermal effects and depends on the sound velocities of the two plasma species. In particular, the length-scale and mass for gravitational collapse are set by the species with the largest sound velocity.

The theoretical framework presented in this paper could be used for a more detailed numerical analysis of the perturbations observed in the {\sc cmb}, incorporating thermal effects in a gauge invariant and covariant fashion.

%%--------------------------------
\section*{Acknowledgments}
%%--------------------------------
This research was supported by the Swedish Research Council 
through the contract No. 621-2004-3217. M.M.\ would like to thank the 
Department of Astrophysics, Radboud University, for their hospitality
during the stay at which this research was initiated and J.M.\ would like to thank the Department of Physics, Ume{\aa} University for returning the favour to complete this work.
\appendix
%-----------------------------------------
\section{Commutation relations}
In this Appendix we summarise the commutation relations that are used throughout this paper and derived in \cite{maartens} and \cite{betschart04}. The relations are linearized about a background in which $\omega_{ab}=\sigma_{ab}=\dot{u}_{a}=0$.  $X$, $X_{a}$, $X_{ab}$ stand for any scalar function, vector or tensor and ${\cal F}$ can be any of those.
\bsub\label{eq::commutations}
\begin{align}
(\D_{a} {\cal F})^{\cdot} &= \D_{a} \dot{{\cal F}} - \tfrac{1}{3} \Theta \D_{a}{\cal F}\ ,\\
(\D^{b} X_{ab})^{\cdot} &= \D^{b} \dot{X}_{ab} - \tfrac{1}{3} \Theta \D^{b} X_{ab}\ ,\\
(\curl X_{ab})^{\cdot}&= \curl\dot{X}_{ab} - \tfrac{1}{3}\Theta \ \curl X_{ab}\ ,\\
\D_{a}\D^{2}X &= \D^{2} \D_{a} X + \tfrac{2}{9}\left(\Theta^{2} - 3\mu \right) \D_{a} X\ ,\\
\D_{a}\D^{2}X^{a} &= \D^{2} \D_{a} X^{a} - \tfrac{2}{9}\left(\Theta^{2} - 3\mu \right) \D_{a} X^{a}\ ,\\
S\ \curl (\D^{2} X^{a}) &= \D^{2}(S\ \curl X^{a})\ . 
\end{align}
\esub

%\bibliography{twotemps.bib}

\end{document}